\documentclass{article}
\usepackage{spconf,amsmath,epsfig}
\usepackage{amsmath,amsthm,amssymb,amsfonts}
\usepackage{multirow}
\usepackage[dvipsnames, svgnames, x11names]{xcolor} 

\let\OLDthebibliography\thebibliography
\renewcommand\thebibliography[1]{
  \OLDthebibliography{#1}
  \setlength{\parskip}{0pt}
  \setlength{\itemsep}{0pt plus 0.3ex}
}

\begin{document}\sloppy

\setlength{\abovedisplayskip}{0pt}

\def\x{{\mathbf x}}
\def\L{{\cal L}}

\title{A real-time blind quality-of-experience assessment metric\\ for HTTP adaptive streaming}
%
\name{Chunyi Li \textsuperscript{1}, May Lim\textsuperscript{2} , Abdelhak Bentaleb \textsuperscript{3}, and Roger Zimmermann\textsuperscript{2} \vspace*{-4mm}}
\address{Shanghai Jiao Tong University\textsuperscript{1}, National University of Singapore\textsuperscript{2}, Concordia University\textsuperscript{3} \vspace*{-4mm}}

\maketitle

\begin{abstract}
In today's Internet, HTTP Adaptive Streaming (HAS) is the mainstream standard
for video streaming, which switches the bitrate of the video content based on an Adaptive BitRate (ABR) algorithm. An effective Quality of Experience (QoE) assessment metric can provide crucial feedback to an ABR algorithm. However, predicting such real-time QoE on the client side is challenging. The QoE prediction requires high consistency with the Human Visual System (HVS), low latency, and blind assessment, which are difficult to realize together. To address this challenge, we analyzed various characteristics of HAS systems and propose a non-uniform sampling metric to reduce time complexity.
Furthermore, we design an effective QoE metric that integrates resolution and rebuffering time as the Quality of Service (QoS), as well as spatiotemporal output from a deep neural network and specific switching events as content information. These reward and penalty features are regressed into quality scores with a Support Vector Regression (SVR) model. Experimental results show that the accuracy of our metric outperforms the mainstream blind QoE metrics by 0.3, and its computing time is only 60\% of the video playback, indicating that the proposed metric is capable of providing real-time guidance to ABR algorithms and improving the overall performance of HAS.
\end{abstract}
\section{Introduction}
\label{sec:intro}

Nowadays, video has become the dominant application on the Internet. According to Cisco's survey \cite{cisco2020cisco}, video services already consume more than 80\% of current Internet capacity and demand is still growing. To meet the challenges posed by the transmission of large volumes of video data, content providers often use HTTP Adaptive Streaming (HAS),
which can adapt to dynamic network conditions and various device resolutions.
HAS delivers media content in small segments over the HTTP/TCP protocol stack, and because of its adoption by leading content providers, HAS has become the dominant delivery method for Video on Demand (VoD) services.


In a HAS client (\emph{i.e.,} the media player), an Adaptive BitRate (ABR) algorithm selects a suitable bitrate level for the download of each video segment. The goal of ABR algorithms~\cite{bentaleb2018survey} is to maximize the user's Quality of Experience (QoE). Therefore, an effective objective QoE metric is crucial in HAS systems. It is either used after a video session ends to evaluate the performance of an ABR algorithm or during playback to guide the ABR algorithm in selecting the most appropriate bitrate demand for the next video segment.


Designing an effective objective QoE metric for bitrate guidance is challenging due to the following three requirements.
($i$) \textbf{High Consistency with HVS}: The QoE metric needs to be consistent with HVS so that it can accurately predict the user's perceived quality of video playback. ($ii$) \textbf{Low Latency}: To provide guidance during transmission, the feedback of the QoE metric should be computed along with the video playback, and not after transmission like in ABR performance evaluation. In this case, the computation time of the QoE metric cannot be longer than the segment's playback time, and hence its complexity should be reduced to ensure real-time feedback. Finally, ($iii$) \textbf{Blind Assessment}: The real-time prediction mentioned above is performed on the client. Hence, this should be a No-Reference (NR) task that uses only the compressed/distorted video available at the client, instead of Full-Reference (FR) or Reduced-Reference (RR) scenarios which use the original uncompressed source video or some of its reference features that have to be separately acquired from the server.
%

\section{Related Work and Contributions}
\label{sec:relatedworks}
Generally speaking, existing QoE metrics that provide guidance for HAS delivery can be classified into three types \cite{C:KSQI}: QoS-based, signal fidelity/content-based, and hybrid metrics.

\begin{figure*}[tbp]
	\centering
	\includegraphics[width=0.98\textwidth]{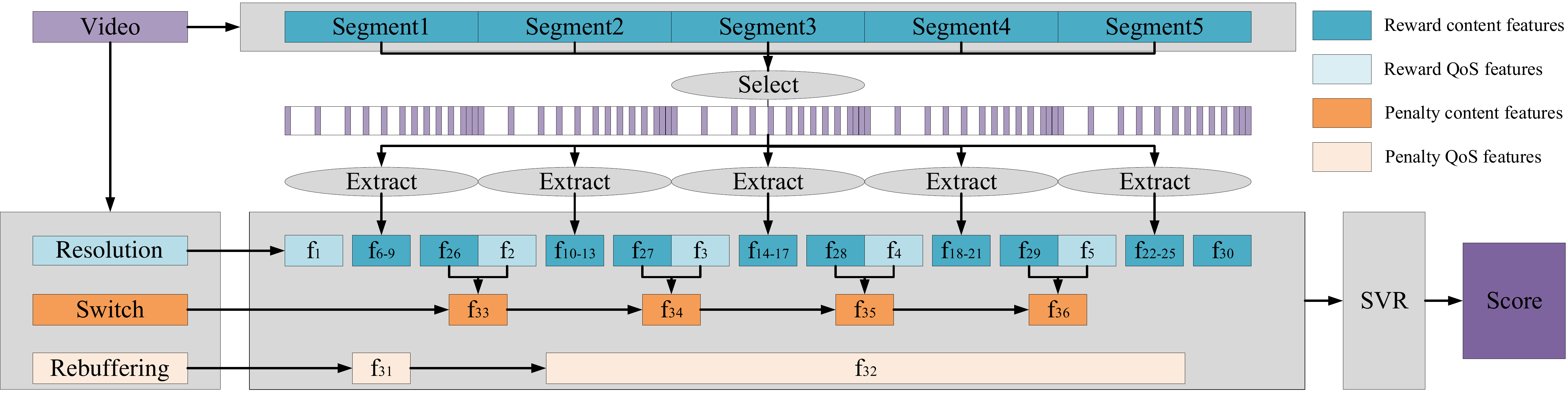}
 \vspace{-3mm}
	\caption{The framework of the proposed method.}
	\label{fig:Framework}
 \vspace*{-5mm}
\end{figure*}

\textbf{QoS-based} metrics \cite{A:FTW,A:Yin2015} have the lowest complexity as their computation typically involves a simple combination of network and/or client-side data, such as video bitrate, startup latency, and rebuffering duration.
However, such metrics tend to show the least consistency with HVS as the video content is not considered in its computation, which has a strong influence on perceptual quality.

\textbf{Content-based} metrics \cite{B:Brisque,B:Viideo} analyze the video content and its signal fidelity using image or video quality assessment (I/VQA) metrics to predict a video's distortion level \cite{Zhai2020PerceptualIQ}. IQA metrics take selected video frames as input and build a time series model \cite{NR:Series} based on the quality of these frames to output a quality score for the video. Since an IQA metric needs to be computed repeatedly for each frame, it may not scale well. Existing VQA metrics also tend to have long run times.
Only a few simplified NR-VQA metrics \cite{B:Rapique, B:FastVQA} are able to provide real-time feedback for ABR, and such simplification leads to some degradation in their performance.

\textbf{Hybrid} metrics \cite{C:Bentaleb2016,C:SQI} combine data from both the QoS and video content. Their QoE score is generally composed of a reward function represented by IQA/VQA metrics and a penalty function based on QoS factors. This allows the metric to reach a balance between time complexity and consistency with HVS, which is promising for its use in bitrate guidance.
Unfortunately, some of the existing hybrid metrics have IQA/VQA kernels that are designed for FR tasks only \cite{C:Bentaleb2016}, while the universal models \cite{C:TV-QoE} tend to show high consistency with HVS only on the FR or RR kernels but low consistency when it comes to NR. Therefore, the performance of such hybrid NR metrics can be further improved, as we show in this work.

As none of the existing metrics above can satisfy the three requirements of Section \ref{sec:intro} altogether, we propose a new QoE assessment metric for HAS with the following contributions. ($i$) We perform a non-uniform sampling scheme since HVS has an increasing focus tendency throughout each segment. Without analyzing too many frames at the start of a segment, gradually increasing the sampling rate can reduce time complexity. ($ii$) We introduce QoS into reward features. As NR functions are not as effective as FR/RR, some QoS information can help our model perform better for blind assessment. ($iii$) We introduce content into penalty features. By analyzing specific frame changes instead of QoS fluctuations, the user's QoE can be better characterized.

\section{Proposed Method}
\label{sec:proposed}

To design a QoE metric that can effectively meet the requirements discussed in Section~\ref{sec:intro}, we identified novel features and adaptations that contribute materially to these requirements. The framework of our blind QoE metric is shown in Fig.~\ref{fig:Framework} and includes three parts: \emph{sampling}, \emph{feature extraction}, and \emph{quality regression}. Taking inspiration from hybrid metrics in Section~\ref{sec:relatedworks}, we extract four types of features: reward/penalty QoS features and reward/penalty content features. For content features, each video segment in the client's buffer is first non-uniformly sampled to select a suitable subset of image frames for feature extraction. The QoS/content features are extracted by ResNet-50, texture analysis, and inter-frame difference, and then regressed through a support vector regression (SVR) model to give a quality score representing the user's current QoE. We discuss the details of each step below.

\vspace*{-1mm}
\subsection{Sampling}
\label{sec:samp}
\vspace*{-1mm}

For real-time QoE assessment, content-based/hybrid metrics analyze only a subset of frames to reduce complexity. 
While traditional metrics sample each segment uniformly, it is generally believed that frames within a segment may not contribute equally to QoE. To study their respective contributions, we run Brisque \cite{B:Brisque}, a widely used NR-VQA metric, on the
Waterloo sQoE \uppercase\expandafter{\romannumeral3} video dataset~\cite{w3}. As each segment is two seconds long 
and an intra-coded frame usually appears every one second~\cite{HAScoding}, we divide a segment into two halves (start/end) {for non-uniform sampling} and represent its $QoE$ as: 
$QoE {\,=\,} {w_s}Qo{E_s} + {w_e}Qo{E_e}$,
where $w_{s}$, $w_{e}$ are weight parameters for the QoE of the start/end of a segment $Qo{E_s}$ and $Qo{E_e}$, respectively. The Spearman Rank-order Correlation Coefficient (SRoCC) is used as the correlation function $\mathcal{S}$ between $QoE$ and the single-stimulus mean opinion scores (MOS) $M$ obtained from the subjective assessment on the dataset: $\mathcal{S}(QoE,M) {\,=\,} {w_s}\mathcal{S}(Qo{E_s}^{f{r_s}},M) + {w_e}\mathcal{S}(Qo{E_e}^{f{r_e}},M)$,
where $fr_s$ and $fr_e$ are the sampled frames from the start/end of a segment. 

To better understand the relationship between sampling rate and correlation performance, we first used eight different sampling rates and three common IQA metrics (Niqe~\cite{B:Niqe}, Piqe~\cite{B:Piqe}, Brisque~\cite{B:Brisque}) to predict QoE. The normalized SRoCC between predicted QoEs and subjective scores show that the correlation factor is logarithmically related to the sampling rate as seen in Fig.~\ref{fig:Log}.
Hence, the sampling scheme can be transformed into the following optimization problem:
\vspace*{-6pt}

\begin{figure}[tbp]
    \centering	\includegraphics[width=0.48\textwidth]{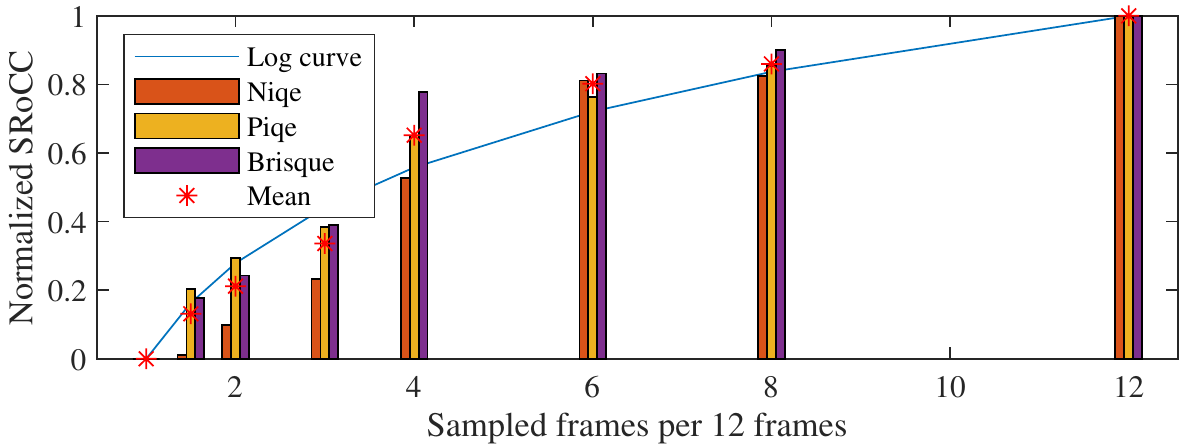}
    \vspace*{-7mm}
	\caption{{Approximate convex logarithmic relationship between sampling rate and the normalized SRoCC.}}
    \label{fig:Log}
    \vspace*{-3mm}
\end{figure}

{\small
\begin{equation}
    \left\{ \begin{array}{l}
fr = f{r_s} + f{r_e}\\
{\mathop{\rm maximize}\nolimits} ({w_s}\log (f{r_s}) + {w_e}\log (f{r_e}))
\end{array} \right.
    \label{equ:convex}
\end{equation}
}
\hspace*{-2mm}
where $fr$ is the total number of sampled frames desired. Via 
Lagrange multiplier, the derivative of the log function in (\ref{equ:convex}) gives $fr$ as proportional to $w$ for the best sampling scheme:
\vspace*{-6pt}

{\small
\begin{equation}
    \frac{{{fr_s}}}{{{fr_e}}} =\frac{{{w_s}}}{{{w_e}}} = \frac{{\mathcal{S}({\mathop{\rm brisque}\nolimits}(seg_s),M)}}{{\mathcal{S}({\mathop{\rm brisque}\nolimits}(seg_e),M)}}
    \label{equ:sampling}
\end{equation}
}
\hspace*{-2mm}
where $seg_s$ and $seg_e$ are the start/end of a segment and QoE is predicted by ${\rm brisque}(\cdot)$. 
The detailed proportion and derivation are attached in the supplementary.

\subsection{Feature Extraction}

We discuss how the four types of features are extracted below.

\textbf{Reward QoS Feature}. QoS refers to the basic network or client-side metrics during a video streaming session.
Among them, video bitrate, quantization parameter (QP), frame rate, and resolution (height/width) have a positive impact on the user's QoE. After studying the correlation performance of these factors, 
we found video height performs well as a reward feature $r_1$ to characterize the perceptual quality. Hence, $r_1={\rm height}(seg)$,
%
%
where $seg$ is a video segment in HAS. 


\textbf{Reward Content Features}. 
After sampling the frames in Section~\ref{sec:samp}, we analyze this series of images which has three types of attainable features: structural, temporal, and chrominance. Structural features in images can reflect perceptual quality very well~\cite{Structural1} 
and are widely used for QoE prediction. Temporal features can be computed independently or from the structural features via a spatial-temporal fusion.
Chrominance features, like structural features, are computed independently from images. However, as the HVS processes visual signals in three channels, computing chrominance features may increase complexity considerably.
To meet the real-time requirement, we include structural features and integrate temporal features with spatial-temporal fusion, while abandoning chrominance features.

The structural features can be divided into spatial and texture features. Spatial features refer to the relative spatial positioning or orientation of different elements in an image. 
An overview of our spatial feature extraction method is shown in Fig.~\ref{fig:ResNet}. ResNet-50 is the backbone of the module, which can represent the spatial correlation between pixels and has proven to be quality-aware \cite{Qaware}. For the $i$-th sampled frame in a segment, we transform it into a gray map $g(i)$ as the input to ResNet-50, 
which extracts four features $L_i={l_{1\sim4}}(i)$ as:
\vspace*{-2pt}
{\small
\begin{equation}
    \left\{ {\begin{array}{l}
    {P(i) = {\rm{ResNet50}}(g(i))}\\
    {[{l_1}(i),{l_2}(i)] = [\max (P(i)),\min (P(i))]}\\
    {[{l_3}(i),{l_4}(i)] = [\rm{avg} {(P(i))},\rm{std} {(P(i))}]}
\end{array}} \right.
    \label{equ:ResNet}
\end{equation}
}
\hspace*{-2mm}
where $P(n)$ is the 
output from ResNet-50 and $i$ is the frame index.
%
%
On a segment level, instead of computing the four spatial features as a global average, we introduce a gated recurrent unit (GRU) \cite{GRUfang} to capture the temporal relation between the spatial features. The spatiotemporal rewards $r_{2\sim5}$ are:
\vspace*{-6pt}

\begin{figure}[btp]
	\centering
	\includegraphics[width=0.48\textwidth]{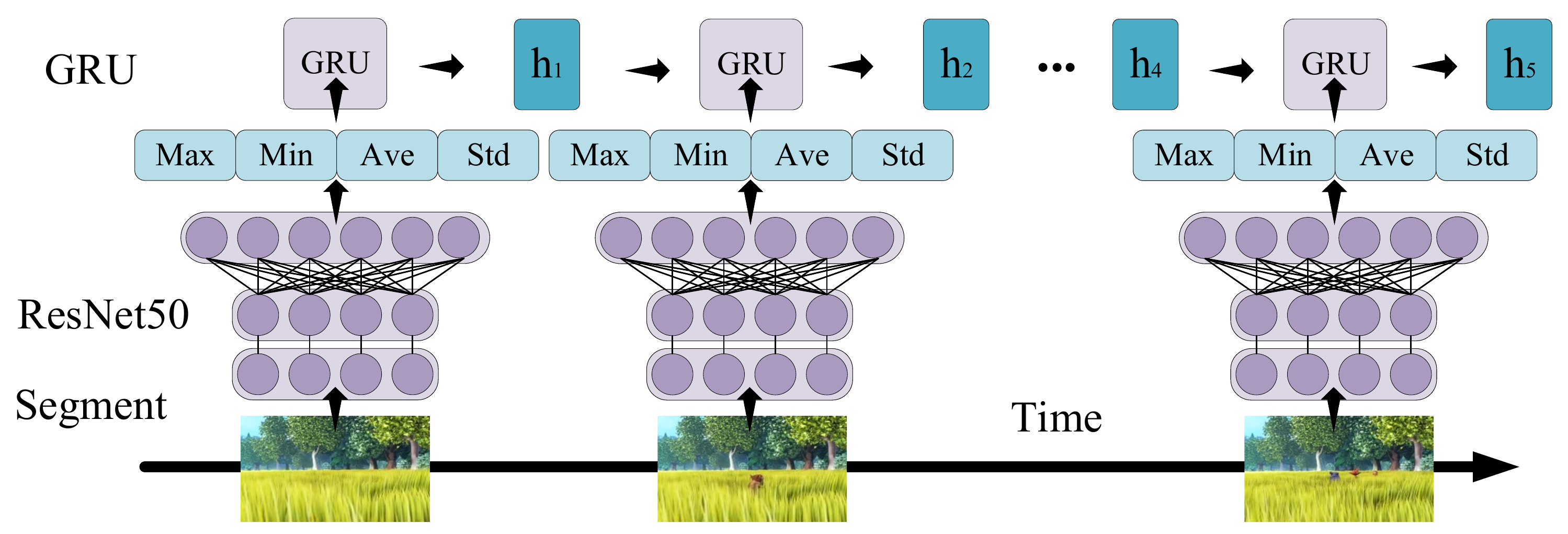}
        \vspace*{-7mm}
	\caption{The spatial feature extraction method.}
	\label{fig:ResNet}
        \vspace*{-3mm}
\end{figure}

{\small
\begin{equation}
    \left\{ {\begin{array}{*{20}{l}}
{{h_i} = GRU\left( {W\cdot{L_i} + b,{\rm{ }}{h_{i - 1}}} \right)}\\
r_{2\sim5}=h_{fr}
\end{array}} \right.
    \label{equ:r2-5}
\end{equation}
}
\hspace*{-2mm}
where $W$ and $b$ are the weights and bias parameters in the
GRU \cite{GRUfang}, while ${h_i}$ is a four-cell meta-array like ${L_i}$ that restores the memory-forgetting mechanism of HVS in the time domain. Thus, these four features combine both spatial and temporal information.

Texture features refer to the surface characteristics of objects within images.
Although operators such as Sobel, Laplace \cite{operator} are commonly used for texture extraction, given the complexity that ResNet-50 has already introduced for spatial features above, a simpler solution is needed here.
%
%
Similar to how MacroBlocks (MBs) are used as { the basic processing unit in video compression~\cite{Patrick4}}, we first compute the average row and column intensity values $Ra_y$, $Ca_x$ from its gray map.
Then we divide the gray map $g(i)$ into multiple 16$\times$16 MBs. For each MB, we calculate the difference between the gray map value and the above-average values in the respective directions (horizontally, vertically, and diagonally). We select the minimum difference above as an MB's texture, and combine them into the texture feature $r_6$ as:
{\small
\begin{equation}
\left\{ {\begin{array}{*{20}{l}}
{Ho{r_j} = \sum {|R{a_y}(i) - {g_{x,y}}(i)|} }\\
{Ve{r_j} = \sum {|C{a_x}(i) - {g_{x,y}}(i)|} }\\
{Di{a_j} = \sum {|0.5(C{a_x}(i) + R{a_y}(i)) - {g_{x,y}}(i)|} }\\
{{r_6} = \sum {\min (Ho{r_j},Ve{r_j},Di{a_j})} }
\end{array}} \right.
\label{equ:r6}
\end{equation}
}
\hspace*{-2mm}
where $Hor$, $Ver$, and $Dia$ are the texture information computed in three directions using the row average $R{a_y}$ and column average $C{a_x}$, while $j$ is the MB index. 


\textbf{Penalty QoS Features}. Among QoS metrics, rebuffering has one of the largest negative impacts on QoE \cite{rebuffering1}. The rebuffering duration and number of rebuffering events are common penalty features used in QoE models~\cite{A:Mok2011, C:Bentaleb2016}.
Besides, players tend to bear an initial buffering (of fewer than two seconds) in exchange for less rebuffering during playback \cite{rebuffering2}. We also note that the negative impact on QoE scales linearly with both rebuffering duration and number of rebuffering events and so we only consider one of them.
Hence, we include the initial buffering duration and average rebuffering duration as penalty features $p_1$ and $p_2$:
{\small
\begin{equation}
    \left\{ {\begin{array}{*{20}{l}}
{{p_1} = {D_1}}\\
{{p_2} = \frac{1}{{T - 1}}\sum\limits_{t = 2}^T {{D_t}} }
\end{array}} \right.
    \label{equ:p1p2}
\end{equation}
}
\hspace*{-3mm}
where $T$ is the total number of segments, $t$ is the segment index, and $D_t$ is the buffering duration of the segment.


\textbf{Penalty Content Features}. 
Research into video QoS analysis \cite{w3} has shown that bitrate switching and rebuffering events can create a poor experience for users, the impact of which depends on several factors, including: ($i$) Switching pattern: It is generally believed \cite{C:Bentaleb2016} that dropping from high to low bitrate creates a more undesirable experience than going from low to high bitrate, and a long rebuffering duration \cite{Stalling} can further amplify the negative effects of such switching event.
($ii$) Video content: For a video where objects are moving slowly with minimal scene changes, the impact of a bitrate switching or rebuffering event is limited, while in an action movie, if the switching/rebuffering occurs during intense motion or when the scene has just changed, it generally leads to a significant drop in user's QoE.
The negative impact of each switching (or rebuffering) event is traditionally calculated using the change in inter-segment bitrate (or rebuffering duration), without considering the video content.
Conversely, we use the efficient FastSSIM \cite{fastSSIM} metric to represent the inter-frame difference in video content between segments.
%
%
The penalty content features $p_3$ combine {all three factors above, namely the rebuffering time, switching level, and content mentioned in supplementary} as shown in (\ref{equ:p3}).
{\small
\begin{equation}
\left\{ {\begin{array}{*{20}{c}}
{swh = {\mathop{\rm ReLU}\nolimits} (bitrate(se{g_t}) - bitrate(se{g_{t + 1}}))}\\
{{p_3} = (1 + \frac{{D_t}}{C_1})(1 + \frac{{swh}}{C_2})/{\mathop{\rm ssim}\nolimits} (se{g_t},se{g_{t + 1}})}
\end{array}} \right.
    \label{equ:p3}
\end{equation}
}
\hspace*{-2.5mm}
where $swh$ is the inter-segment bitrate differential mapped to a ReLU function as it can characterize bitrate decline in the switching pattern mentioned above, while $C_1$ and $C_2$ are constants for normalization. The FastSSIM function ${\rm ssim}(\cdot)$ is performed on the last sampled frame in segment $seg_t$ and the first sampled frame in $seg_{t+1}$. 

\vspace*{-2mm}
\subsection{Overall QoE Regression}

After the above operations, we finally obtain two reward feature groups $r_1$ and $r_{2 \sim 6}$, and two penalty feature groups $p_{1,2}$ and $p_{3}$ to represent the positive and negative impact on user's QoE, respectively. The feature groups are mapped to features $f_{1 \sim 36}$ as shown in Table~\ref{tab:feature}. Using these features extracted from QoS and video content information, a quality prediction model is constructed via support vector regression (SVR) to generate the overall QoE score as shown in Fig.~\ref{fig:Framework}.
\begin{table}[]
\caption{The list of feature groups and their definitions.}
\small
\hspace*{-3mm}
\begin{tabular}{|c|c|c|c|}
\hline
Feature Group                   & Component & Origin & Description                            \\ \hline
Reward QoS                      & $f_{1\sim5}$      & $r_1$     & Resolution             \\ \hline
\multirow{2}{*}{Reward Content} & $f_{6\sim25}$     & $r_{2\sim5}$   & Spatial-Temporal \\ \cline{2-4} 
                                & $f_{26\sim30}$    & $r_6$     & Texture       \\ \hline
Penalty QoS                     & $f_{31\sim32}$    & $p_{1\sim2}$   & Rebuffering                       \\ \hline
Penalty Content                 & $f_{33\sim36}$    & $p_3$     & 
\begin{tabular}{@{}c@{}}Switching, Rebuffering \\ (content-aware)\end{tabular}
\\ \hline
\end{tabular}
\label{tab:feature}
\vspace{-4mm}
\end{table}
%
As suggested in prior studies \cite{w3}, we use the most recent five segments in the playback for prediction in the SVR model.
The SVR model is implemented using LIBSVM \cite{libsvm} with {a radial basis function (RBF) kernel for feature fusion\cite{Patrick2}}.

\begin{table*}[t]
	\centering
        \small
        \vspace*{-2mm}
	\caption{Performance results on the Waterloo sQoE \uppercase\expandafter{\romannumeral3} and LIVE Netflix \uppercase\expandafter{\romannumeral2} datasets.}
	\label{performance}
\begin{tabular}{|c|c|c|cccc|cccc|}
\hline
\multirow{2}{*}{Type}    & \multirow{2}{*}{Subtype} & \multirow{2}{*}{Method} & \multicolumn{4}{c|}{Waterloo sQoE \uppercase\expandafter{\romannumeral3}}                                                                                                             & \multicolumn{4}{c|}{LIVE Netflix \uppercase\expandafter{\romannumeral2}}                                                                                                              \\ \cline{4-11} 
                         &                          &                         & SRoCC & KRoCC & \multicolumn{1}{c|}{PLCC ↑} & Time ↓ & SRoCC & KRoCC & \multicolumn{1}{c|}{PLCC ↑} & Time ↓\\ \hline
\multirow{9}{*}{Content} & \multirow{5}{*}{IQA}     & Brisque5\cite{B:Brisque}               & 0.4959                        & 0.3468                        & \multicolumn{1}{c|}{0.4690}                       & 1.944                        & 0.3146                        & 0.2182                        & \multicolumn{1}{c|}{0.2009}                       & 1.614                        \\
                         &                          & Brisque10\cite{B:Brisque}               & 0.4842                        & 0.3394                        & \multicolumn{1}{c|}{0.4600}                       & 0.965                        & 0.2704                        & 0.1880                         & \multicolumn{1}{c|}{0.1344}                       & 1.276                        \\
                         &                          & Brisque20\cite{B:Brisque}               & 0.4547                        & 0.3146                        & \multicolumn{1}{c|}{0.4478}                       & 0.498                        & 0.2543                        & 0.1794                        & \multicolumn{1}{c|}{0.1517}                       & 0.455                        \\
                         &                          & Niqe10\cite{B:Niqe}                  & 0.4021                        & 0.2732                        & \multicolumn{1}{c|}{0.4488}                       & 1.027                        & 0.6538                        & 0.4761                        & \multicolumn{1}{c|}{0.6684}                       & 0.529                        \\
                         &                          & Piqe10\cite{B:Piqe}                  & 0.4232                        & 0.2807                        & \multicolumn{1}{c|}{0.4349}                       & 1.128                        & 0.6746                        & 0.4898                        & \multicolumn{1}{c|}{0.6871}                       & 0.541                        \\ \cline{2-11} 
                         & \multirow{4}{*}{VQA}     & VIIDEO\cite{B:Viideo}                  & 0.3946                        & 0.2651                        & \multicolumn{1}{c|}{0.4903}                       & 8.047                        & 0.2843                        & 0.1885                        & \multicolumn{1}{c|}{0.3228}                       & 5.624                        \\
                         &                          & V-BLIINDS\cite{B:Vblind}               & 0.7389                        & 0.5456                        & \multicolumn{1}{c|}{0.7244}                       & 45.625                       & 0.7510                         & 0.5755                        & \multicolumn{1}{c|}{0.7653}                       & 42.755                       \\
                         &                          & Resnet50\cite{B:Resnet}                & 0.5707                        & 0.4113                        & \multicolumn{1}{c|}{0.5635}                       & 0.381                        & 0.4278                        & 0.2995                        & \multicolumn{1}{c|}{0.4317}                       & 0.307                        \\
                         &                          & FAST-VQA\cite{B:FastVQA}                & 0.7391                        & 0.5500                        & \multicolumn{1}{c|}{0.7710}                       & 0.246                        & 0.5137                        & 0.3644                        & \multicolumn{1}{c|}{0.5748}                       & 0.232                        \\ \hline
\multirow{5}{*}{QoS}     & \multirow{2}{*}{Client}    & FTW\cite{A:FTW}                     & 0.1835                        & 0.1337                        & \multicolumn{1}{c|}{0.3229}                       & 0.001                        & 0.0804                        & 0.0858                        & \multicolumn{1}{c|}{0.0648}                       & 0.001                        \\
                         &                          & MoK2011\cite{A:Mok2011}                 & 0.1687                        & 0.1294                        & \multicolumn{1}{c|}{0.2156}                       & 0.001                        & 0.0795                        & 0.0650                        & \multicolumn{1}{c|}{0.0874}                       & 0.001                        \\ \cline{2-11} 
                         & \multirow{3}{*}{Network} & Liu2012\cite{A:Liu2012}                 & 0.2529                        & 0.1717                        & \multicolumn{1}{c|}{0.2424}                       & 0.001                        & 0.6633                        & 0.4684                        & \multicolumn{1}{c|}{0.6366}                       & 0.001                        \\
                         &                          & Xue2014\cite{A:Xue2014}                 & 0.3412                        & 0.2245                        & \multicolumn{1}{c|}{0.3081}                       & 0.003                        & 0.5830                        & 0.4123                        & \multicolumn{1}{c|}{0.4961}                       & 0.003                        \\
                         &                          & Yin2015\cite{A:Yin2015}                 & 0.1458                        & 0.0932                        & \multicolumn{1}{c|}{0.3232}                       & 0.007                        & 0.0804                        & 0.0616                        & \multicolumn{1}{c|}{0.0648}                       & 0.007                        \\ \hline
\multirow{5}{*}{Hybrid} & \multirow{5}{*}{Mix}     & SQI\cite{C:SQI}                     & 0.1515                        & 0.1100                        & \multicolumn{1}{c|}{0.2225}                       & 0.501                        & 0.7347                        & 0.5298                        & \multicolumn{1}{c|}{0.6329}                       & 0.458                        \\
                         &                          & TV-QoE\cite{C:TV-QoE}                  & 0.5068                        & 0.3565                        & \multicolumn{1}{c|}{0.4667}                       & 0.524                        & 0.6686                        & 0.4136                        & \multicolumn{1}{c|}{0.5109}                       & 0.482                        \\
                         &                          & Bentaleb2016\cite{C:Bentaleb2016}            & 0.1979                        & 0.1387                        & \multicolumn{1}{c|}{0.3405}                       & 0.498                        & 0.4454                        & 0.2982                        & \multicolumn{1}{c|}{0.4530}                       & 0.456                        \\
                         &                          & KSQI\cite{C:KSQI}                    & 0.5285                        & 0.3875                        & \multicolumn{1}{c|}{0.5268}                       & 0.505                        & 0.7394                        & 0.5492                        & \multicolumn{1}{c|}{0.7315}                       & 0.462                        \\ \cline{3-11} 
                         &                          & Proposed                & \textbf{0.8627}               & \textbf{0.6871}               & \multicolumn{1}{c|}{\textbf{0.8824}}              & 0.606                        & \textbf{0.7739}               & \textbf{0.5914}               & \multicolumn{1}{c|}{\textbf{0.7898}}              & 0.691                        \\ \hline
\end{tabular}
\vspace*{-5mm}
\end{table*}

\section{Performance Evaluation}

\subsection{Experiment Setup}
\label{experimentsetup}

The proposed metric is validated on the Waterloo sQoE \uppercase\expandafter{\romannumeral3} \cite{w3} and the LIVE Netflix \uppercase\expandafter{\romannumeral2} \cite{LIVE2} datasets, which contain various subjectively-rated videos of diverse content types and video codecs, and streamed over various network conditions and ABR algorithms. The subjective study was done using dual-task single-stimulus (SS) experiments with user ratings provided as ground truth. The dataset is split randomly in an 80/20 ratio for training/testing while ensuring the same video content falls into the same set \cite{B:Rapique}.
The partitioning and evaluation process is repeated 1,000 times for a fair comparison, and the average result is reported as the final performance.

%
%
We evaluate our metric in the following ways: First, to evaluate its consistency with HVS, we use three common correlation functions, namely SRoCC, Kendall rank-order correlation coefficient (KRoCC), and Pearson linear correlation coefficient (PLCC), to measure how well our metric correlates with the subjective scores. Second, to evaluate its latency, we measure the computation time of our metric on a laptop with an i7-8750H CPU, which is a common client specification for HAS services \cite{cisco2020cisco}. 
Third, since this is a blind assessment scenario, we compare our metric to 18 mainstream blind QoE assessment metrics as baselines. 
For model training, we left the parameters of ResNet-50 unchanged (which has been pre-trained for image classification on ImageNet \cite{imagenet}), and update the other parameters using the Adam optimizer \cite{adam}. The other learning-based methods are trained with similar mechanisms for a fair comparison. In the IQA models, we evaluated three different sampling rates by uniformly sampling every 5, 10, and 20 frames; in the hybrid models, as their original SSIM feature is not attainable for the NR task, we uniformly sample every 20 frames and apply the widely used Brisque~\cite{B:Brisque} metric in their content-based features.

\subsection{Experiment Results and Discussion}

\begin{figure}[tbp]
	\centering
	\includegraphics[width=0.45\textwidth]{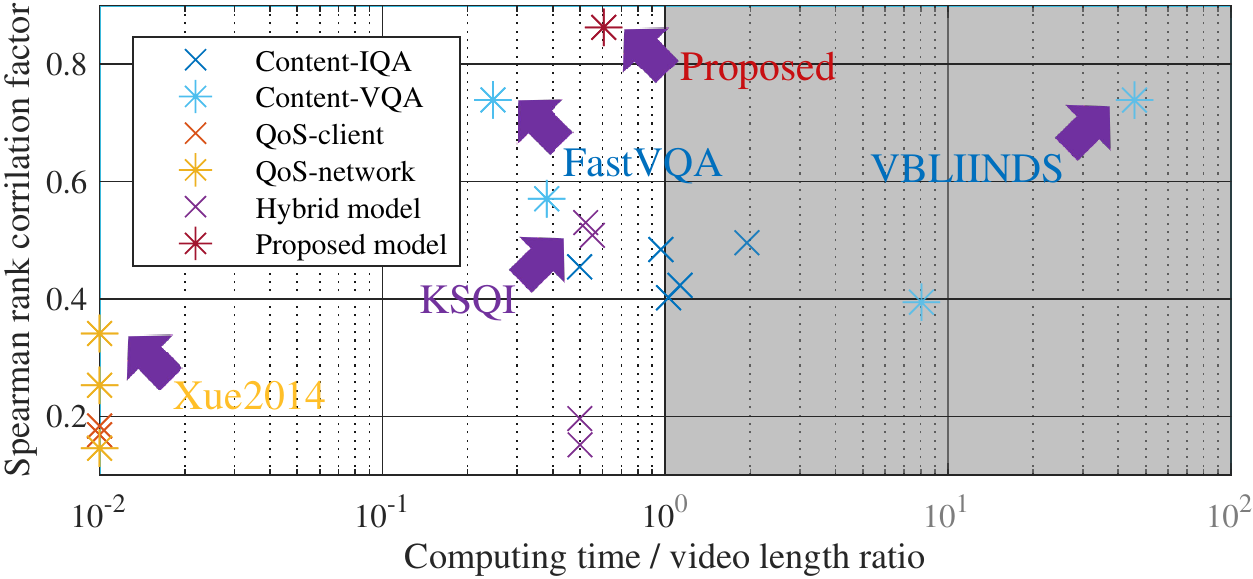}
        \vspace*{-4mm}
	\caption{SRoCC and time ratio performance of the baseline and proposed methods. Metrics with time ratio $<$ 1 (white panel) can realize real-time QoE prediction for bitrate guidance.}
	\label{fig:vis}
        \vspace*{-5mm}
\end{figure}

Table \ref{performance} and Fig.~\ref{fig:vis} show the performance results of the baseline and proposed methods, 
where a larger correlation factor indicates a higher consistency with the HVS.
Computation time is calculated as a ratio of the video duration, where a smaller ratio indicates lower complexity and a value below 1 is needed to meet the real-time requirement.
Among the content-based metrics, V-BLIINDS~\cite{B:Vblind} has the best prediction performance but requires a computation time that is more than 40$\times$ longer than the video itself, resulting in its inability to provide real-time feedback for bitrate guidance.
The QoS-based metrics have much faster computation times of less than 1\% of the video duration, but the predicted QoE has a relatively poor correlation with subjective scores.
%
Most of the hybrid models achieve a better balance between consistency and latency. However, they are generally designed for FR features but their prediction performance becomes less ideal when switched to NR.
%
%
Results show that our metric outperforms all QoS-based and hybrid metrics in all three correlation measures for both datasets (with a gain of up to 0.35 against other hybrid metrics), and outperforms all content-based metrics in SRoCC and PLCC for both datasets while keeping the computation time ratio below 1.
%
{In terms of global QoE metric performance factors\cite{Patrick1}, our Area Under the ROC Curve (AUC) value is about 0.09 ahead of SOTA HAS QoE metric and outperforms all current metrics for correct classification.}
Hence, our metric can overcome the absence of reference information in blind assessment scenarios and still provide effective QoE predictions that have high consistency with HVS (with an average SRoCC of 0.81) and sufficiently low latency (of about 65\% of the video playback duration).

\vspace*{-1mm}
\subsection{Ablation Study}
\vspace*{-1mm}

To validate the contributions of our sampling method and the different feature types, we also conduct an ablation study and its results are shown in Table \ref{abandon}. The factors are specified as: (1) Non-uniform sampling, (2) Reward QoS features, (3) Reward content features (both spatial and textural), (4) Penalty QoS features, and (5) Penalty content features.
The results show that removing any single factor leads to performance degradation, which confirms that they all contribute to the performance results in Table \ref{performance}. 
{The time cost for each group can ensure a real-time assessment.}
Ablation results also show that QoS feature groups (2)(4) are more effective than the content feature groups (3)(5), and a combination of them achieves desirable performance,
while the sampling method (1) further enhances the model's performance with little extra time cost.

\begin{table}[t]
\vspace*{-3mm}
\centering
\small
\caption{Performance results of abandoning different features on the Waterloo sQoE \uppercase\expandafter{\romannumeral3} dataset.}
\label{abandon}
\begin{tabular}{|c|ccc|c|}
\hline
Abandoned & SRoCC & KRoCC & PLCC & Time \\ \hline
None & \textbf{0.8627} & \textbf{0.6871} & \textbf{0.8824} & 0.606 \\
(1) & 0.8514 & 0.6716 & 0.8694 & 0.605 \\
(2) & 0.4521 & 0.3270 & 0.4761 & 0.606 \\
{(3) $r_2 \sim r_5$} & {0.7662} & {0.5822} & {0.8194} & {0.381} \\
{(3) $r_6$} & {0.8468} & {0.6680} & {0.8637} & {0.441} \\
(3) All & 0.7313 & 0.5478 & 0.7490 & 0.216 \\
(4) & 0.8319 & 0.6544 & 0.8691 & 0.606 \\
(5) & 0.8434 & 0.6767 & 0.8639 & 0.390 \\
(2)(4) & 0.3204 & 0.2214 & 0.4006 & 0.605 \\
(3)(5) & 0.7010 & 0.5215 & 0.7073 & 0.001 \\ \hline
\end{tabular}
\vspace*{-4mm}
\end{table}


\vspace*{-1mm}
\section{Conclusions}
\vspace*{-2mm}

In this study, we target the challenge of measuring perceptual quality in HAS clients for bitrate guidance.
A blind QoE assessment metric is proposed to provide QoE feedback of high consistency with HVS, at low latency, and without the use of reference information. Specifically, we map QoS and video content information to both reward and penalty features and include a non-uniform sampling mechanism to identify relevant frames for analysis.
Experiments show that the proposed metric achieves the best results in {two databases across three correlation measures}, which suggests strong consistency with HVS, while meeting latency and blind assessment requirements.
This metric is suited to perform real-time QoE assessment on the client side, which can help improve the overall resource utilization of HAS services. 


\end{document}